\documentstyle[prb,aps]{revtex}

\begin{document}

\draft

\preprint{FIMAT-2/95}

\title{Frenkel excitons in random systems with correlated Gaussian
disorder}

\author{F.\ Dom\'{\i}nguez-Adame}

\address{Departamento de F\'{\i}sica de Materiales,
Facultad de F\'{\i}sicas, Universidad Complutense,
E-28040 Madrid, Spain}

\maketitle

\begin{abstract}

Optical absorption spectra of Frenkel excitons in random one-dimensional
systems are presented.  Two models of inhomogeneous broadening, arising
from a Gaussian distribution of on-site energies, are considered.  In
one case the on-site energies are uncorrelated variables whereas in the
second model the on-site energies are pairwise correlated (dimers).  We
observe a red shift and a broadening of the absorption line on
increasing the width of the Gaussian distribution.  In the two cases we
find that the shift is the same, within our numerical accuracy, whereas
the broadening is larger when dimers are introduced.  The increase of
the width of the Gaussian distribution leads to larger differences
between uncorrelated and correlated disordered models.  We suggest that
this higher broadening is due to stronger scattering effects from
dimers.

\end{abstract}


\narrowtext

\section{Introduction}

Recently, several researchers have shown that structural {\em correlated
disorder} has profound effects in random systems and produces a variety
of unexpected phenomena.  It is by now well known that a band of
delocalized electrons appears in tight-binding Hamiltonians with
correlated diagonal and/or off-diagonal elements.
\cite{Flores1,Dunlap,Wu,Phillips,Flores2} A more dramatic occurrence of
electron delocalization arises in continuous models with dimer
impurities as well as in disordered semiconductor superlattices with
dimer quantum wells, where there exist infinitely many bands of
delocalized states. \cite{PRBKP,Diez1,Diez2} What is most important,
delocalization by correlation is not restricted to electrons, as
suggested by the occurrence of delocalized vibrations in classical
random chains with paired disorder. \cite{nozotro,Datta} Furthermore,
time-domain analysis of Frenkel excitons in systems with traps randomly
placed in an otherwise perfect lattice shows that the coherent quantum
transport is also affected by pairing of these traps, but trapping is
less effective as compared to systems with the same fraction of unpaired
traps. \cite{PRBF}

Roughly speaking, correlated disorder seems to retain some of those
special features characterizing periodic systems; in other words, there
exists a competition betweeen short-range correlation and long-range
disorder leading, for instance, to delocalization of electrons and
phonons or to the occurrence of a major contribution of the slowly
decaying modes in exciton trapping processes, as mentioned above.
However, in this paper we point out that this conjecture is not always
valid.  In particular, we consider optical absorption of Frenkel
excitons in unpaired as well as in paired disordered models, focusing
our attention on inhomogeneous broadening due to a Gaussian distribution
of on-site energies in a one-dimensional lattice.  We show that
inhomogeneous broadening is enhanced when structural correlations arise
since exciton scattering from dimers is stronger as compared to
scattering from unpaired sites.

\section{Model}

We consider $N$ optically active centers in a regular lattice with
spacing unity.  For our present purposes, we neglect all thermal degrees
of freedom (electron-phonon coupling and local lattice distortions).
Therefore, the effective Hamiltonian for the Frenkel-exciton problem
can be written in the tight-binding form with nearest-neighbor
interactions as follows (we use units such that $\hbar=1$)
\begin{equation}
{\cal H}= \sum_{k}\>\epsilon_{k} a_{k}^{\dag}a_{k} +
T\sum_{k}\>(a_{k}^{\dag}a_{k+1}+a_{k+1}^{\dag}a_{k}),
\label{perfectH}
\end{equation}
where $a_{k}$ and $a_{k}^{\dag}$ are Bose operators creating or
annihilating an electronic excitation of on-site energy $\epsilon_k$.
Here $T$ is the nearest-neighbor coupling, which is assumed to be
constant in the whole lattice.  On-site energies are subject to diagonal
disorder representing inhomogeneous broadening, for which a Gaussian
distribution is the proper theoretical approximation.  In what follows
we consider two different models, namely uncorrelated and correlated
disordered systems.  This enable us to separate the effects merely due
to optical absorption in one-dimension from those which manifest the
peculiarities of the correlation between random parameters.  The first
model can be drawn from the distribution \cite{Huber,Wager}
\begin{equation}
P(\epsilon_k,\overline{\epsilon},\sigma)={1\over \sqrt{2\pi\sigma^2}}\,
\exp\left[-\,\frac{(\epsilon_k-\overline{\epsilon})^2}{2\sigma^2}\right],
\label{P}
\end{equation}
with average $\overline{\epsilon}$ and width $\sigma$.  It is important
to stress that neighbor on-site energies are uncorrelated random
variables.  On the other side, to build up our correlated disordered
model, we chose $\epsilon_{2k-1}$ according to the Gaussian distribution
given in (\ref{P}) and then take $\epsilon_{2k}=\epsilon_{2k-1}$.  This
step is repeated at every odd site of the lattice, hence leading to a
set of paired correlated on-site energies (dimers).

Having presented our model we now describe the method we have used to
calculate the absorption spectra.  The line shape $I(E)$ of optical
absorption in which a single exciton is created after pulse excitation
in a lattice with $N$ sites can be obtained as follows \cite{Huber}
\begin{equation}
I(E)=-\,{2\over \pi N} \int_0^\infty\> dt\, e^{-\alpha t} \sin (Et)
\,\mbox{\rm Im} \left( \sum_k\> G_k(t) \right),
\label{line}
\end{equation}
where the factor $\exp (-\alpha t)$ takes into account the broadening
due to the Lorentzian instrumental resolution function of half width
$\alpha$. Hence the resulting spectra are the convolution of the
inhomogeneous broadening due to Gaussian disorder and the instrumental
resolution function. The correlation functions $G_{k}(t)$ obeys the
equation of motion
\begin{equation}
i{d\over dt} G_{k}(t) = \sum_{j}\> H_{kj} G_{j}(t),
\label{motion}
\end{equation}
with $H_{kj}=\epsilon_k \delta_{kj}+(1-\delta_{kj})T$ with
$j=k-1,k,k+1$, and initial condition $G_{k}(0)=1$.  Therefore, the
computation of optical spectra reduces to solving the discrete
Schr\"odinger-like equation (\ref{motion}) using standard numerical
techniques.

\section{Numerical results and discussions}

We have solved the equation of motion (\ref{motion}) for chains of
$N=2\,000$ sites using an implicit integration scheme.  In order to
minimize end effects, spatial periodic boundary conditions are
introduced.  Energy will be measured in units of $T$ whereas time will
be expressed in units of $T^{-1}$.  To make contact with experiments, we
note that $|T|$ is proportional to the exciton bandwidth in the perfect
lattice, so that energy and time scales can be deduced from experimental
data.  Since we are mainly interested in the comparison between
uncorrelated and correlated disordered models rather than in the effects
of the different parameters on the optical absorption process, we will
fix the values of $\overline{\epsilon}$ and $T$ focusing our attention
on the width $\sigma$.  In particular, we have set
$\overline{\epsilon}=4$ and $T=-1$ henceforth as representative values.
The width of the instrumental resolution was $\alpha=1/4$.  The
distribution width $\sigma$ ranged from $0$ up to $0.5$.

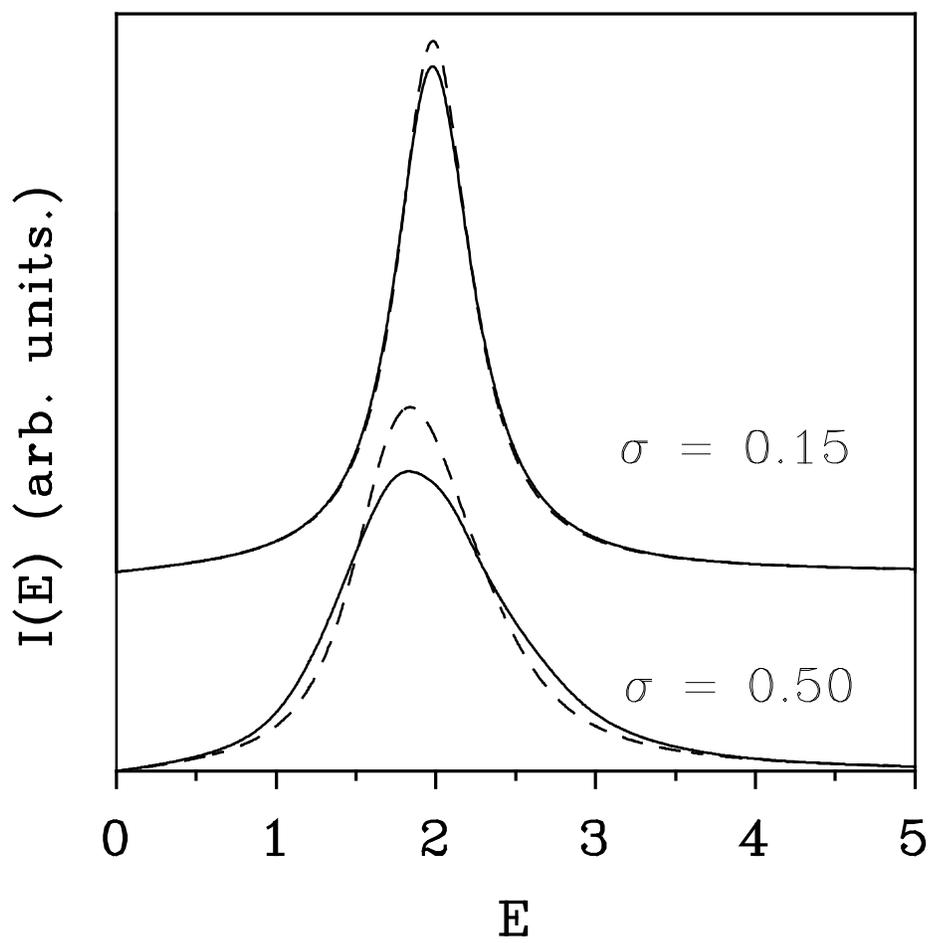
\begin{figure}
\caption{Absorption spectra for one-dimensional random lattices with
Gaussian distribution of correlated (solid lines) and uncorrelated
(dashed lines) on-site energies of with $\sigma=0.15$ and $\sigma=0.50$.
Vertical scale is the same in all cases.}
\label{fig1}
\end{figure}

In the absence of inhomogeneous broadening ($\sigma=0$), the absorption
line shape is a Lorentzian function centered at
$E=\overline{\epsilon}+2T$, which with our choice of parameters is
$E=2$.  The full width at half maximum (FWHM) is $2\alpha$.  For $\sigma
\neq 0$ a broadening of this main line is observed accompanied by a
shift of its position to lower energies on increasing $\sigma$, as shown
in Fig.~\ref{fig1} for correlated as well as uncorrelated disorder.
Unlike optical spectra in random binary systems, where $\epsilon_k$
takes only two different unpaired \cite{Huber} or paired \cite{PRBOPT}
values, there are no signatures of satellites appearing in the
high-energy region of the spectra, at least in the range of parameters
we have considered.  Only a long high-energy tail is observed on
increasing $\sigma$.

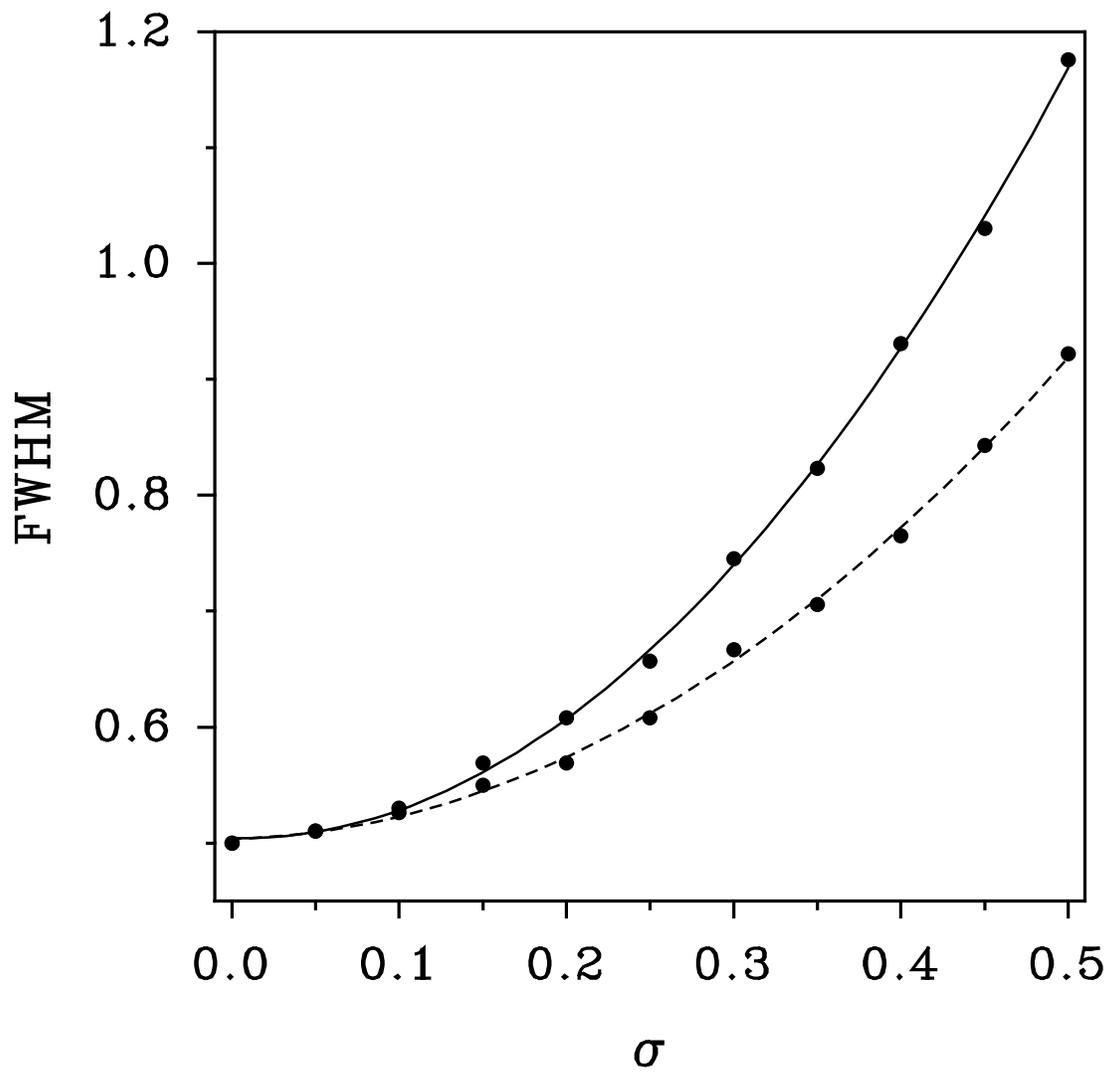
\begin{figure}
\caption{Full width at half maximum (FWHM) as a function of $\sigma$ for
correlated (solid lines) and uncorrelated (dashed lines) disordered
systems.}
\label{fig2}
\end{figure}

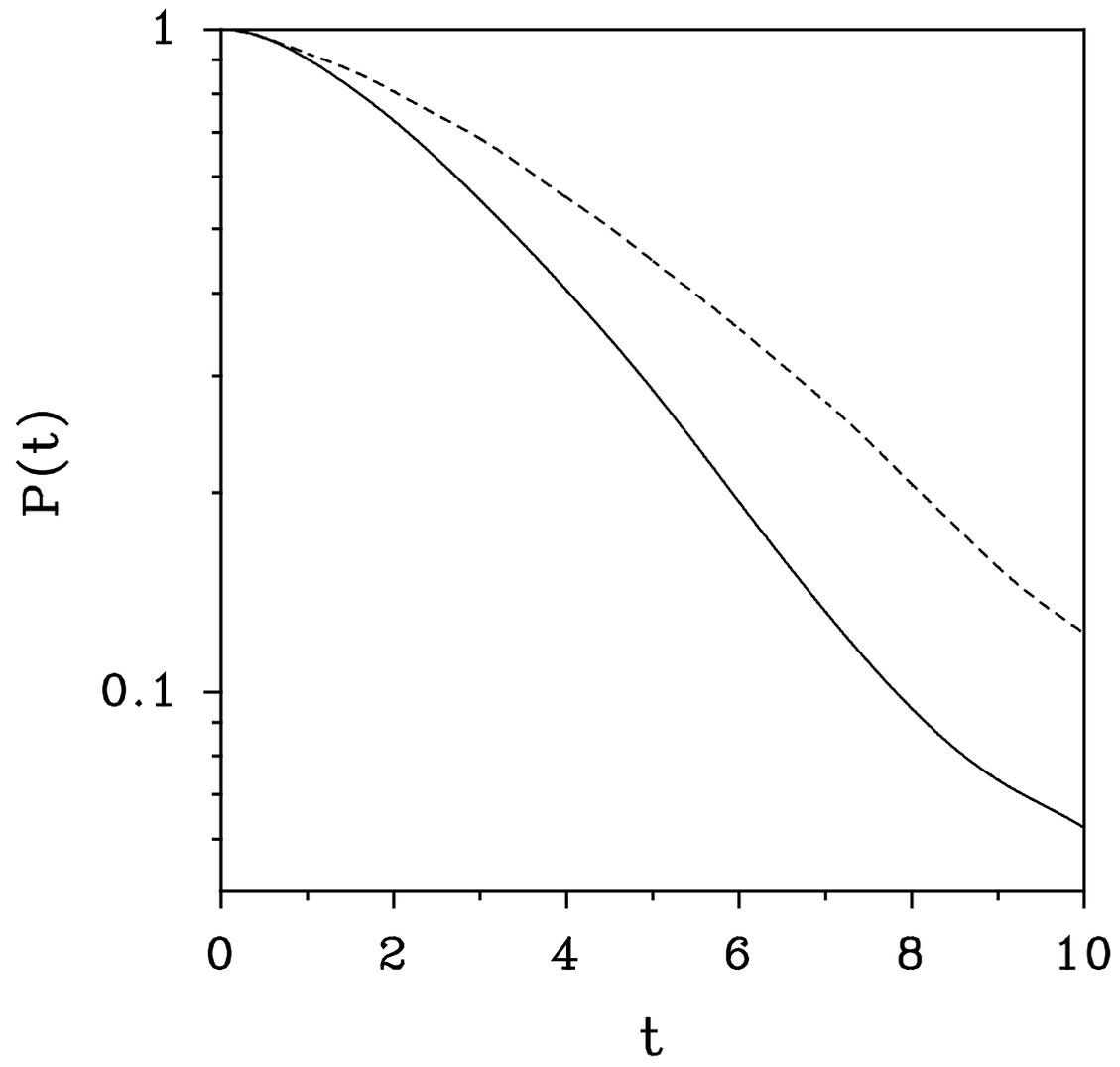
\begin{figure}
\caption{Probability of finding an exciton in the $q=0$ mode at time $t$
for correlated (solid lines) and uncorrelated (dashed lines) disordered
systems with $\sigma=0.50$.}
\label{fig3}
\end{figure}

{}From Fig.~\ref{fig1} several conclusions can be drawn.  First, the shift
to lower energies increases with increasing $\sigma$ in both models.  A
similar shift is also found using the coherent potential approximation
(CPA) but the CPA fails to predict its value correctly, bringing up a
smaller shift than that observed in numerical simulations. \cite{Huber}
Second, this {\em red} shift is the same in both models for a given
value of $\sigma$, within our numerical accuracy.  Hence, we are led to
the conclusion that the effects of correlation cannot be deduced from
this shift.  Third, and more important from an experimental viewpoint,
there are substantial differences regarding the width of the optical
spectra.  In all cases we have studied we found that the FWHM is larger
for correlated inhomogeneous broadening, and that the difference
increases on increasing $\sigma$.  This result is shown in
Fig.~\ref{fig2}, where it is seen that the FWHM is $2\alpha=0.5$ close
to $\sigma=0$, as expected, but increases more pronouncedly when
correlation is present.  This result is somewhat unexpected since it
indicates that correlated Gaussian disorder disturbs exciton dynamics
much more than uncorrelated disorder.  This is to be compared with
previous results on random systems with traps appearing pairs in an
otherwise perfect lattice, where pairing causes less disruption of
Frenkel-exciton dynamics as compared to those systems without this
constraint, \cite{PRBF} as mentioned in the Introduction.  In
particular, the depletion of the $q=0$ exciton mode is slower when traps
are paired.  However, our present results suggest that in the case of
inhomogeneous broadening the behavior is just the opposite, i.e.,
correlations create centers causing larger scattering effects on
excitons.  To validate or discard this suggestion we have evaluated the
probability of finding an exciton in the $q=0$ mode at time $t$ after
pulse excitation, obtained as follows \cite{Huber2}
\begin{equation}
P(t)={1\over N^2}\,\left|\sum_k G_k(t)\right|^2.
\label{q0}
\end{equation}
Figure~\ref{fig3} shows the results for $\sigma=0.50$, although we have
found similar behavior in all cases.  From this figure we can see that
$P(t)$ decays faster when correlations are present.  Since in our system
there is no trapping, the $q=0$ mode is only depleted by disorder, thus
indicating that scattering by dimers is more important than scattering
by uncorrelated single sites.  In this sense correlated systems are {\em
more} disordered that uncorrelated ones, thus explaining the larger
inhomogeneous broadening in the former systems.

Finally, let us comment that the CPA predicts a much smaller FWHM than
that found by numerical integration in the case of inhomogeneous
broadening in uncorrelated disordered lattices. \cite{Huber}  Therefore,
the CPA failure is even more dramatic in the case of correlated
disorder, strongly suggesting that future theoretical works should rely
on different grounds.  Probably, a {\em renormalization} of the lattice
considering the dimer defect as a single entity, as much as in the line
we have recently proposed in the case of trapping of classical excitons
in correlated disordered systems, \cite{Classical} could be a good
starting theoretical scenario to fully account for optical spectra.

\section{Conclusions}

In summary, we have studied the effects of inhomogeneous broadening on
the absorption spectrum corresponding to the Frenkel-exciton Hamiltonian
for random systems.  Two different models have been considered; in both
cases broadening arises from a Gaussian distribution of on-site
energies.  In uncorrelated disordered systems on-site energies are
chosen according to the Gaussian distribution at every site, whereas in
correlated disordered systems this selection is made only at odd sites
and even sites take the value of the preceding site.  Therefore, the
correlation length is of the order of the lattice spacing.  Our results
show that the presence of structural correlations in random systems can
be then readily determined from the analysis of the absorption spectra
of these samples.  By comparing the obtained spectra in both models we
found an identical red shift of the absorption line while inhomogeneous
broadening is more pronounced whenever correlations are present in the
lattice.  We have realized that the $q=0$ mode is depleted faster in
systems with correlated disorder, thus indicating a stronger scattering
effects from dimers.  From a theoretical viewpoint, we have pointed out
that CPA failures are even more noteworthy in the case of correlated
disordered systems.  Hence, the necessity of a new theoretical framework
to deal with random systems with structural correlations becomes a very
appealing task.

\acknowledgments

The author thanks A.\ S\'{a}nchez and E.\ Maci\'a for illuminating
conversations.  This work is supported by UCM under project No.\
PR161/93-4811.


\end{document}